\documentclass[a4paper,11pt]{article}
\usepackage{pos}

\newcommand{\bqa}{\mbox{\boldmath $q$}_{1}}
\newcommand{\bqb}{\mbox{\boldmath $q$}_{2}}

\title{
Inclusive production of $f_2(1270)$ tensor mesons at the LHC via gluon-gluon fusion in the $k_t$-factorization approach}
\ShortTitle{Inclusive production of $f_2(1270)$ tensor mesons at the LHC via gluon-gluon fusion}

\author*[\dag]{A. Szczurek}
\notes{\note{Also at \textit{College of Natural Sciences, 
Institute of Physics, University of Rzesz{\'o}w, 
Pigonia 1, PL-35310 Rzesz{\'o}w, Poland}.}}
\author{P. Lebiedowicz}
\affiliation{Institute of Nuclear Physics Polish Academy of Sciences,\\ 
Radzikowskiego 152, PL-31342 Krak{\'o}w, Poland}

\emailAdd{Antoni.Szczurek@ifj.edu.pl}
\emailAdd{Piotr.Lebiedowicz@ifj.edu.pl}

\abstract{
The cross section for inclusive production of $f_2(1270)$ meson
is calculated. We include both the mechanism of gluon-gluon fusion
as well as the $\pi \pi$ final-state rescattering. 
The contribution of the gluon-gluon fusion is calculated within
the $k_t$-factorization approach with modern unintegrated gluon
distribution functions (UGDFs).
Some parameters for the $g^* g^* \to f_2$ vertex are extracted from 
the $\gamma \gamma \to f_2(1270) \to \pi \pi$ reactions.
The results strongly depend on the parametrization of the 
$g^* g^* \to f_2(1270)$ form factor.
Results of our model are compared to the ALICE preliminary data.
The gluon-gluon fusion does not explain low-$p_t$ data but could
be the dominant mechanism at somewhat larger meson transverse
momenta.
By adjusting some parameters the pion-pion rescattering can explain
the low-$p_t$ region. 
}

\FullConference{%
  40th International Conference on High Energy Physics - ICHEP2020\\
  July 28 - August 6, 2020\\
  Prague, Czech Republic (virtual meeting)
}

\begin{document}
\maketitle

\section{Introduction}
\vspace{-0.2cm}

The mechanism of $f_2(1270)$ meson production in proton-proton
collisions is not well understood.
For instance in event generators $f_2(1270)$ is not produced
in a primary fragmentation process but occurs only in decays. 
The $f_2(1270)$ is also difficult to observe experimentally.
The dominant decay channel is $f_2(1270) \to \pi^+ \pi^-$.
Only STAR \cite{Adams:2003cc} and ALICE \cite{Lee:thesis} undertook
experimental efforts.

The gluon-gluon fusion was already considered in the past
\cite{FillionGourdeau:2007ee}. Recently \cite{LS2020} we discussed
this mechanism more carefully.

\begin{figure}[!ht]
\begin{center}
\includegraphics[width=0.3\textwidth]{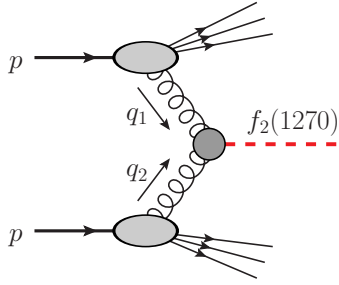}
\caption{\label{fig:diagram_gg_f2}
\small
General diagram for inclusive $f_2(1270)$ production
via gluon-gluon fusion in $pp$ collisions.}
\end{center}
\end{figure}

In figure~\ref{fig:diagram_gg_f2} we show a Feynman diagram for the
$f_2(1270)$ meson production via 
gluon-gluon fusion in proton-proton collisions.

\vspace{-0.6cm}
We applied $k_t$-factorization approach
used by our group for $\chi_c$ quarkonium production
\cite{Cisek:2017gno,Babiarz:2020jkh},
for $\eta_c(1S,2S)$ production 
\cite{Baranov:2019joi,Babiarz:2019mag},
and also for $f_{0}(980)$ production \cite{Lebiedowicz:2020bwo}.
The colour singlet gluon-gluon fusion is similar to
photon-photon fusion discussed e.g. in \cite{Ewerz:2013kda}.
There two tensor structures corresponding to 
$\Gamma^{(0)}$ helicity-0 and $\Gamma^{(2)}$ helicity-2
couplings were found and their strength was
determined from the comparison to the Belle data \cite{Dai:2014zta}
for the $\gamma \gamma \to \pi \pi$ reactions.
In \cite{LS2020} we also used tensorial structures 
for $\gamma^{*} \gamma^{*} \to f_2(1270)$ vertex
from \cite{Pascalutsa:2012pr}. 

\vspace{-0.2cm}
\section{Model calculations}
\vspace{-0.2cm}

In \cite{Ewerz:2013kda} the $f_{2} \gamma \gamma$ vertex 
for `on-shell' $f_{2}$ meson
and real photons was considered;
see (3.39) of \cite{Ewerz:2013kda} and the discussion in section 5.3 there.
In \cite{LS2020} we were interested in 
$\gamma^*(Q_1^2) \gamma^*(Q_2^2) \to f_2(1270)$ process. 
Thus to describe the dependence on photon virtualities
we introduced the vertex form factors 
$F^{(0)}(Q_1^2,Q_2^2)$ and $F^{(2)}(Q_1^2,Q_2^2)$.

Then the $\gamma^* \gamma^* \to f_2(1270)$ vertex,
including the form factors $F^{(\Lambda)}$ (for $\Lambda = 0, 2$)
were parametrized as
\begin{eqnarray}
\Gamma_{\mu\nu\kappa\lambda}^{(f_2 \gamma \gamma)}(q_1,q_2) 
= 2 a_{f_2 \gamma\gamma}\, 
\Gamma_{\mu\nu\kappa\lambda}^{(0)}(q_1,q_2)\,
F^{(0)}(Q_1^2,Q_2^2)-b_{f_2 \gamma\gamma}\, 
\Gamma_{\mu\nu\kappa\lambda}^{(2)}(q_1,q_2) \,
F^{(2)}(Q_1^2,Q_2^2)  \,,
\label{EMN_vertex}
\end{eqnarray}
with two rank-four tensor functions 
$\Gamma^{(0)}$
and 
$\Gamma^{(2)}$
(see (3.18)--(3.22) of \cite{Ewerz:2013kda}).

To obtain $a_{f_{2} \gamma \gamma}$ 
and $b_{f_{2} \gamma \gamma}$ in (\ref{EMN_vertex})
we used the experimental value
of the radiative decay width 
\begin{eqnarray}
&&\Gamma(f_{2} \to \gamma \gamma) = (2.93 \pm 0.40) \,\mathrm{keV}\,,\nonumber \\
&&\mathrm{helicity\; zero \;contribution} \approx 9 \% \;\mathrm{of}\;\Gamma(f_{2} \to \gamma \gamma)\,,
\label{PDG_values}
\end{eqnarray}
as quoted for the preferred solution III 
in Table~3 of \cite{Dai:2014zta}.
Using the decay rate from (5.28) of \cite{Ewerz:2013kda}
\begin{equation}
\Gamma(f_{2} \to \gamma \gamma) =
\frac{m_{f_{2}}}{80 \pi} 
\left( \frac{1}{6} m_{f_{2}}^{6} |a_{f_{2} \gamma \gamma}|^{2} + m_{f_{2}}^{2} |b_{f_{2} \gamma \gamma}|^{2} \right)
\label{decay_rate}
\end{equation}
and assuming $a_{f_{2} \gamma \gamma} > 0$
and $b_{f_{2} \gamma \gamma} >0$, we find
$a_{f_{2} \gamma \gamma}=
\alpha_{\rm em}\, \times\,
1.17\;\mathrm{GeV^{-3}}$ and $b_{f_{2} \gamma \gamma}=
\alpha_{\rm em}\, \times\,
2.46\;\mathrm{GeV^{-1}}$,
where $\alpha_{\rm em} = e^{2}/(4 \pi) \simeq 1/137$.

It was shown in Refs.~\cite{Poppe:1986dq,Schuler:1997yw,Pascalutsa:2012pr} 
that the most general amplitude for the process
$\gamma^\ast (q_1, \lambda_1) + \gamma^\ast(q_2, \lambda_2) 
\to f_{2}(\Lambda)$, 
describing the transition from two  
virtual photons to a tensor meson $f_{2}$ ($J^{PC} = 2^{++}$)
with the mass $m_{f_{2}}$ and 
helicity $\Lambda = \pm 2, \pm 1, 0$, 
involves five independent structures. 

In the formalism in \cite{Pascalutsa:2012pr} 
the $\gamma^* \gamma^* \to f_2(1270)$ vertex 
was parameterized as
\begin{eqnarray}
&&\Gamma_{\mu\nu\kappa\lambda}^{(f_2 \gamma \gamma)}(q_1,q_2) 
= 4 \pi \alpha_{\rm em}
\left\{
 \left[ R_{\mu \kappa} (q_1, q_2) R_{\nu \lambda} (q_1, q_2) 
+ \frac{s}{8 X} \, R_{\mu \nu}(q_1, q_2) 
(q_1 - q_2)_\kappa \, (q_1 - q_2)_\lambda \right]
\right. 
\nonumber \\
&&\quad \quad \left. 
\times \frac{\nu}{m_{f_{2}}} T^{(2)}(Q_1^2, Q_2^2)
\right. 
\nonumber \\
&&\quad \quad \left. + R_{\nu \kappa}(q_1, q_2) (q_1 - q_2)_\lambda  
\left( q_{1 \mu} + \frac{Q_1^2}{\nu} q_{2 \mu} \right) 
\frac{1}{m_{f_{2}}} 
T^{(1)}(Q_1^2, Q_2^2) 
\right. 
\nonumber \\
&&\quad \quad \left. + R_{\mu \kappa}(q_1, q_2) (q_2 - q_1)_\lambda   
\left( q_{2 \nu} + \frac{Q_2^2}{\nu} q_{1 \nu} \right) 
\frac{1}{m_{f_{2}}} 
T^{(1)}(Q_2^2, Q_1^2)
\right. 
\nonumber \\ 
&&\quad \quad \left. + R_{\mu \nu}(q_1, q_2) (q_1 - q_2)_\kappa \, (q_1 - q_2)_\lambda \, \frac{1}{m_{f_{2}}} 
T^{(0, {\rm T})}(Q_1^2, Q_2^2) 
\right. 
\nonumber \\
&&\quad \quad \left. + \left( q_{1 \mu} + \frac{Q_1^2}{\nu} q_{2 \mu} \right) 
\left( q_{2 \nu} + \frac{Q_2^2}{\nu} q_{1 \nu} \right)  
(q_1 - q_2)_\kappa  (q_1 - q_2)_\lambda 
\frac{1}{m_{f_{2}}^3}
T^{(0, {\rm L})}(Q_1^2, Q_2^2)
\right\},
\label{PPV_vertex}
\end{eqnarray}
where photons with four-momenta $q_{1}$ and $q_{2}$ have
virtualities,
$Q_{1}^{2}=-q_{1}^{2}$ and $Q_{2}^{2}=-q_{2}^{2}$,
$s = (q_{1}+q_{2})^{2} = 2 \nu - Q_1^2 - Q_2^2$, 
$X = \nu^{2} - q_1^2 q_2^2$,
$\nu = (q_{1} \cdot q_{2})$, and
\begin{equation}
R_{\mu \nu}(q_{1},q_{2}) = -g_{\mu \nu} + \frac{1}{X}
\left[
\nu \left( q_{1 \mu} q_{2 \nu} + q_{2 \mu} q_{1 \nu} \right)
- q_1^2 q_{2 \mu} q_{2 \nu} - q_2^2 q_{1 \mu} q_{1 \nu}
\right]\,.
\end{equation}
In (\ref{PPV_vertex}) $T^{(\Lambda)}(Q_1^2, Q_2^2)$ 
are the $\gamma^* \gamma^* \to f_2(1270)$ 
transition form factors for $f_2(1270)$ helicity $\Lambda$. 
For the case of helicity zero, 
there are 2 form factors depending on whether both photons are 
transverse (${\rm T}$) or longitudinal (${\rm L}$).
We express the transition form factors as
\begin{eqnarray}
T^{(\Lambda)}(Q_1^2, Q_2^2) = F^{(\Lambda)}(Q_1^2, Q_2^2) \, T^{(\Lambda)}(0,0) \,.
\label{TFF_aux}
\end{eqnarray}
In the limit $Q_{1,2}^{2} \to 0$ 
only $T^{(0,{\rm T})}$ and $T^{(2)}$ contribute
and their values at $Q_{1,2}^{2} \to 0$ determine
the two-photon decay width of $f_{2}(1270)$ meson.

Comparing two approaches,
EMN and PPV,
given by 
(\ref{EMN_vertex}) and (\ref{PPV_vertex})--(\ref{TFF_aux}), respectively,
at both real photons ($Q_{1}^{2} = Q_{2}^{2} = 0$) 
and at $\sqrt{s} = m_{f_{2}}$
we found the correspondence 
\begin{eqnarray}
&&4 \pi \alpha_{\rm em} \, T^{(0,{\rm T})}(0,0) =  
- a_{f_{2} \gamma \gamma} \, \frac{m_{f_{2}}^{3}}{2}\,,\\
&&4 \pi \alpha_{\rm em} \, T^{(2)}(0,0) = 
- b_{f_{2} \gamma \gamma} \, 2 m_{f_{2}} \,.
\label{PPV_parameters}
\end{eqnarray}
%


The $g^* g^* \to f_{2}(1270)$ coupling entering in the matrix element squared
was obtained from that for the $\gamma^* \gamma^* \to f_{2}(1270)$ coupling
by the following replacement:
\begin{equation}
\alpha_{\rm{em}}^2 \to \alpha_{\rm s}^2  \,
\frac{1}{4 N_c (N_c^2 - 1)} \,
\frac{1}{(<e_q^2>)^2} \,.
\label{replacement}
\end{equation}
Above $(<e_{q}^2>)^2 = 25/162$ for 
the $\frac{1}{\sqrt{2}} \left(u \bar u + d \bar d \right)$
flavour structure assumed for $f_{2}(1270)$.

The running of strong coupling constants was included.
In numerical calculations the renormalization scale was taken in the form:
\begin{equation}
\alpha_{\rm s}^2 \to 
\alpha_{\rm s}(\max{\{m_{T}^2,\bqa^2\}})\,
\alpha_{\rm s}(\max{\{m_{T}^2,\bqb^2\}})\,.
\label{alpha_s}
\end{equation}
The Shirkov-Solovtsov prescription \cite{Shirkov:1997wi} is used to 
extrapolate down to small renormalization scales 
relevant for the $f_2(1270)$ production for the ALICE kinematics.


Because $f_2(1270)$ is extended, finite size object one can expect 
in addition a form factor $F(Q_{1}^{2}, Q_{2}^{2})$ 
associated with the gluon virtualities for
the $g^* g^* \to f_2$ vertex.
In \cite{LS2020} the form factor was parametrized 
in four different ways; see (2.13)--(2.16) of \cite{LS2020}.
%
%

\vspace{-0.2cm}
\section{Comparison to ALICE data}
\vspace{-0.2cm}

In \cite{LS2020} we showed our results to the ALICE data
presented in a PhD thesis ~\cite{Lee:thesis}.
To convert to the number of $f_2(1270)$ mesons per event 
we used the relation:
\begin{equation}
\frac{d N}{d p_t} = \frac{1}{\sigma_{\rm inel}} \frac{d \sigma}{d p_t} \,.
\label{dN_dpt}
\end{equation}
The inelastic cross section for $\sqrt{s} = 7$~TeV 
%
%
was obtained by the TOTEM \cite{Antchev:2013gaa} 
and ATLAS \cite{Aad:2014dca} collaborations. 
In our calculations we took $\sigma_{{\rm inel}} = 72.5$~mb.

\begin{figure}[!ht]
\includegraphics[width=0.495\textwidth]{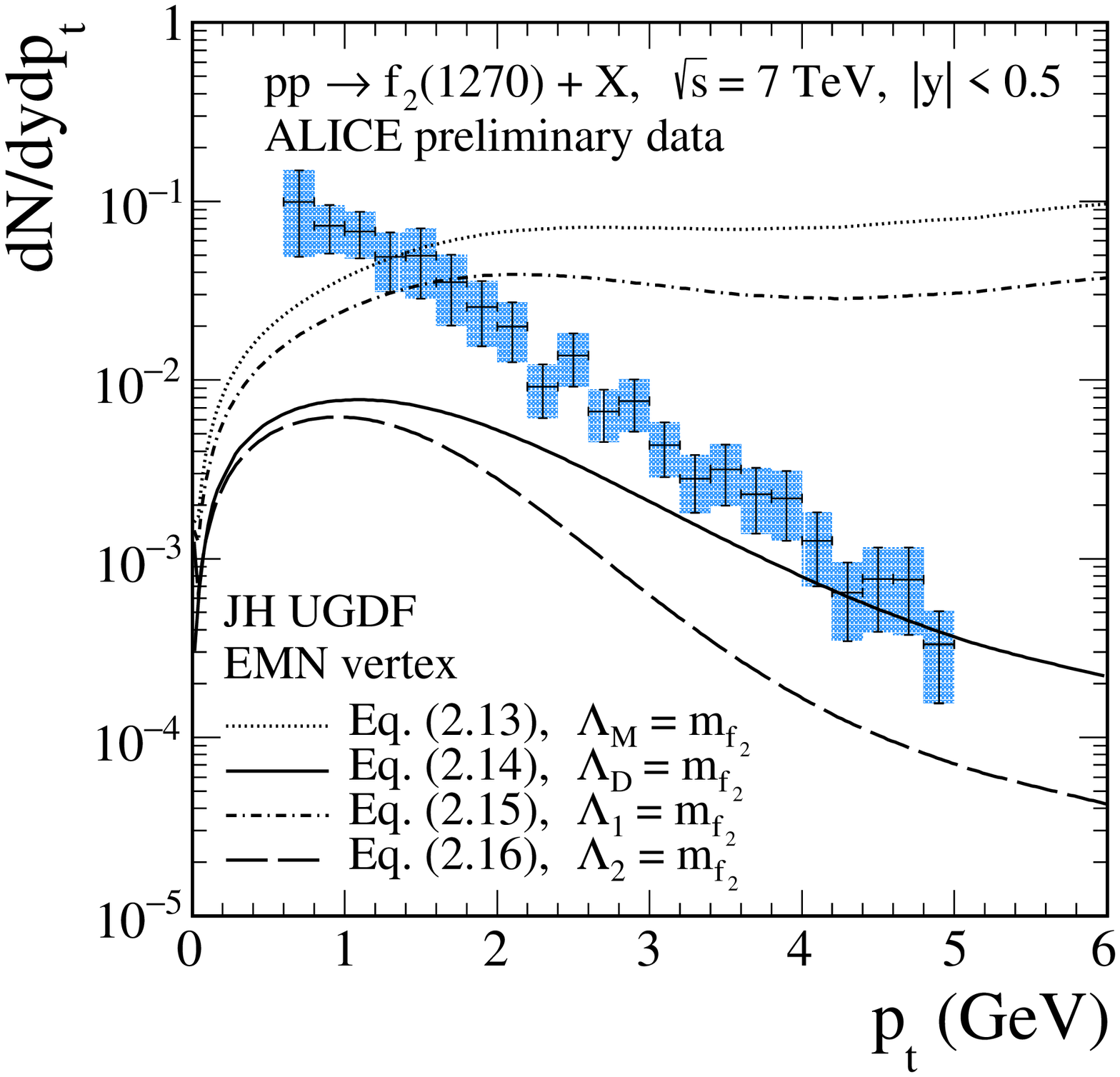}
\includegraphics[width=0.495\textwidth]{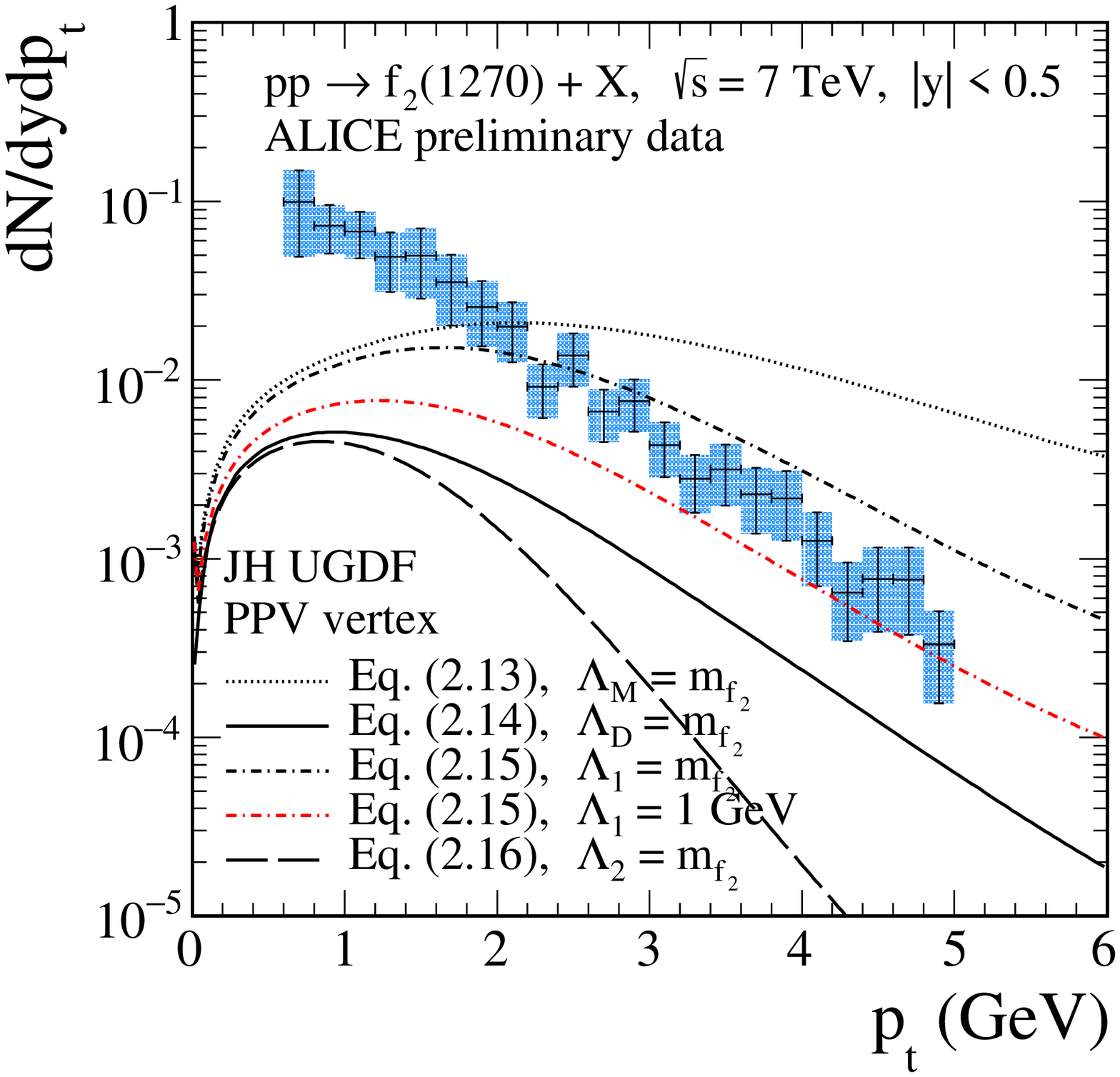}
\caption{\label{fig:dN_dpt_ff}
\small
The $f_{2}(1270)$ meson transverse momentum distributions at
$\sqrt{s}=7$~TeV and $|{\rm y}|<0.5$.
The preliminary ALICE data from \cite{Lee:thesis} 
are shown for comparison.
The results for the EMN (left panel)
and PPV (right panel) $g^* g^* \to f_2(1270)$ vertex
for different $F(Q_1^2,Q_2^2)$ form factor (2.13)--(2.16) from \cite{LS2020} 
are shown.
In this calculation the JH UGDF was used.}
\end{figure}

In figure~\ref{fig:dN_dpt_ff} we present 
the $f_2(1270)$ meson transverse momentum distributions at
$\sqrt{s}=7$~TeV and $|{\rm y}|<0.5$
together with the preliminary ALICE data.
Here, for the color-singlet gluon-gluon fusion mechanism,
we used JH UGDF from \cite{Hautmann:2013tba}.
We compare results for the monopole
and dipole form factors for two different $g^{*}g^{*} \to f_{2}$ vertices.

In the formalism of \cite{Pascalutsa:2012pr} 
[see the PPV vertex (\ref{PPV_vertex})]
there is no interference between 
$\Lambda = 0, {\rm{T}}$ and $\Lambda = 2$ terms while 
the naive use of the formalism from \cite{Ewerz:2013kda} 
[see the EMN vertex (\ref{EMN_vertex})]
generates some interference effects.
Different couplings 
lead to different shapes of the transverse momentum distributions;
see \cite{LS2020} for details and more results.

\begin{figure}[!ht]
\includegraphics[width=0.495\textwidth]{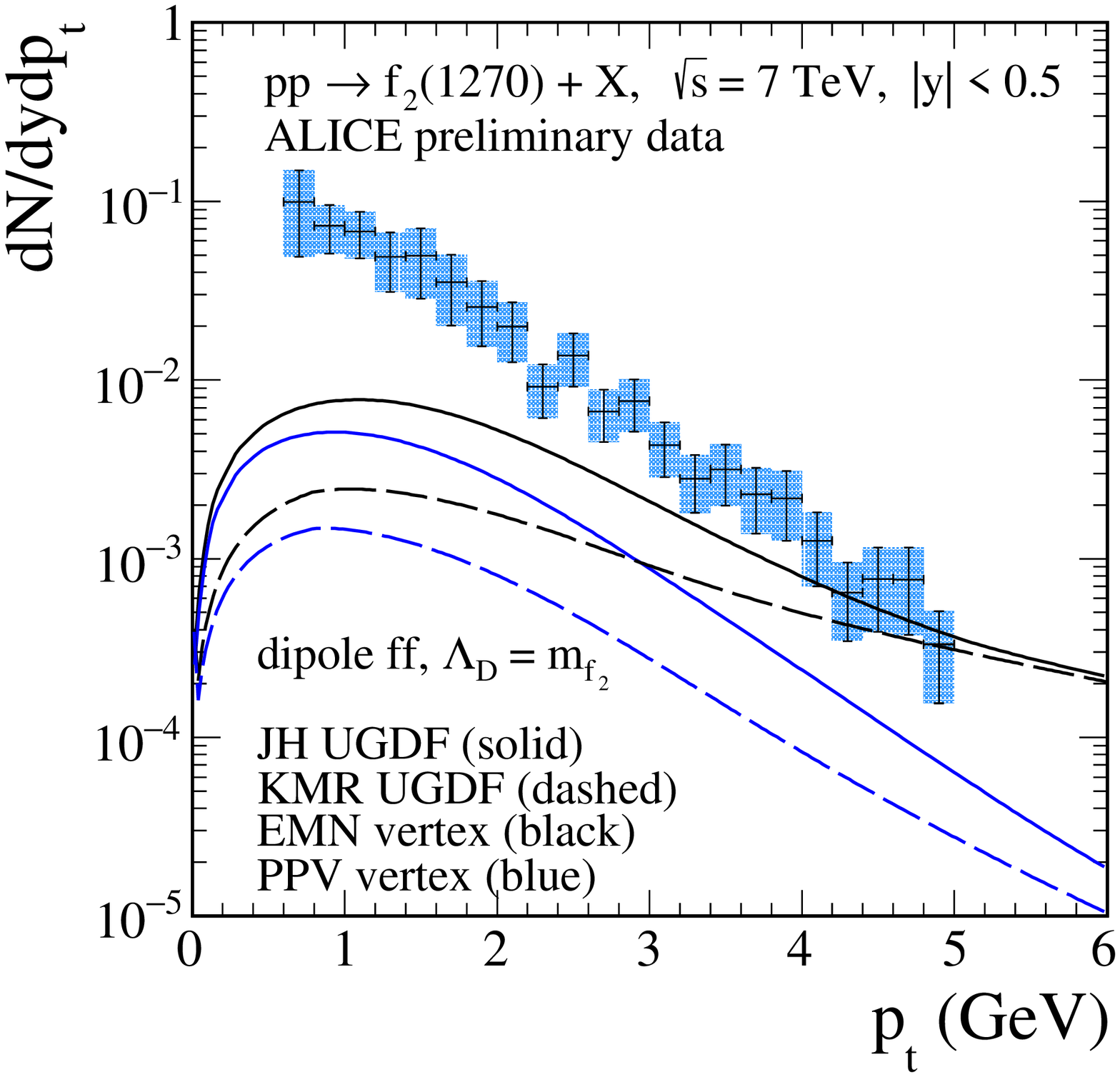}
\includegraphics[width=0.495\textwidth]{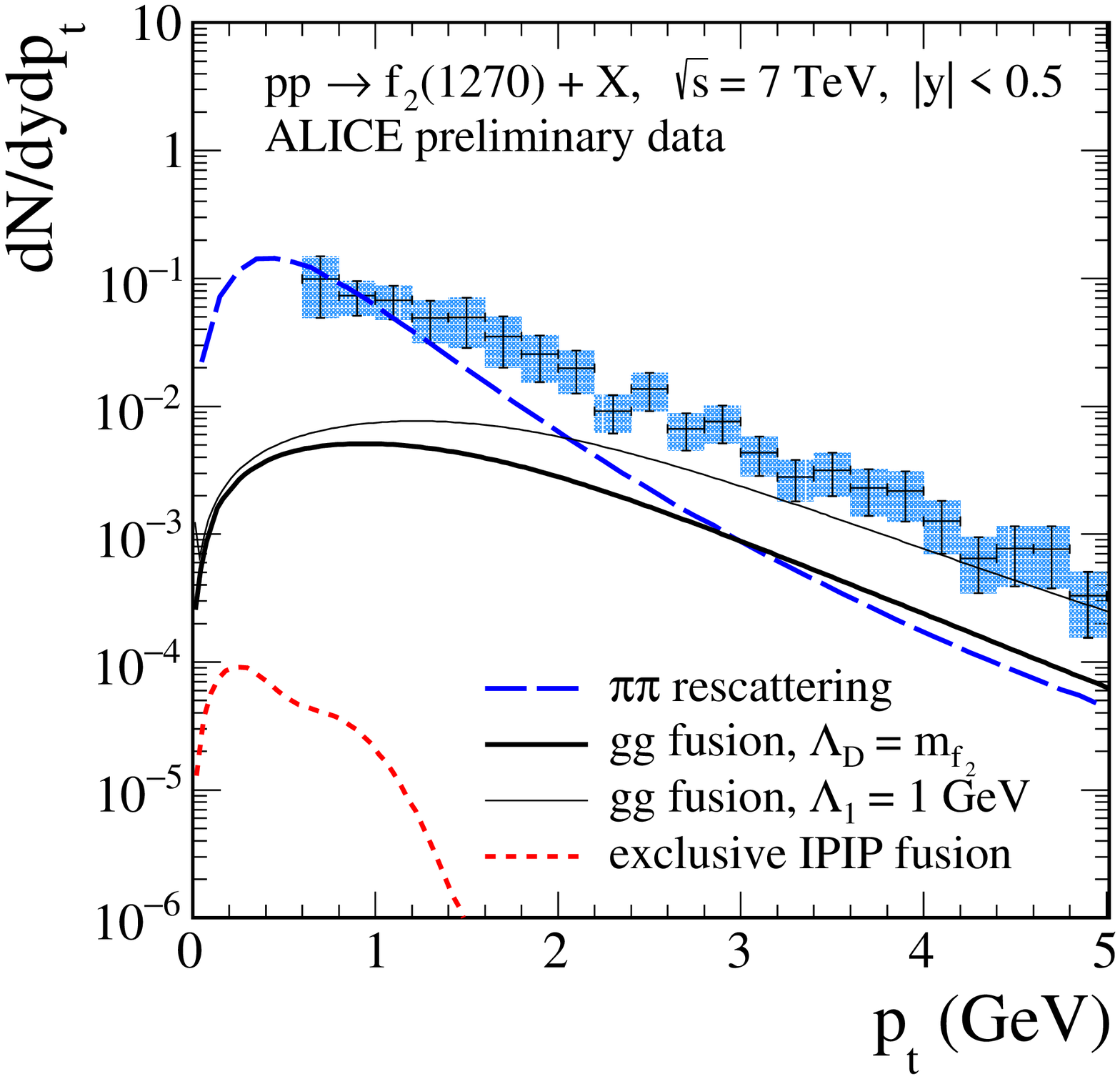}
\caption{\label{fig:GJR}
\small
In the left panel results for two different UGDFs, 
JH (solid lines) and KMR (dashed lines), 
together with the preliminary ALICE data from \cite{Lee:thesis}
are shown. 
In the right panel results for the $\pi \pi$ rescattering mechanism 
(long-dashed line), for the $gg$-fusion (JH UGDF) mechanism (solid lines), 
and for the pomeron-pomeron fusion mechanism (dotted line) are shown.
The dotted (red) line corresponds to the Born-level result 
for the exclusive $pp \to pp f_2(1270)$ reaction via pomeron-pomeron fusion; see \cite{Lebiedowicz:2016ioh,Lebiedowicz:2019por}.
}
\end{figure}

In the left panel of figure~\ref{fig:GJR} we show results 
for the KMR UGDF.
Here we use a glue constructed 
according to the prescription
initiated in \cite{Kimber:2001sc} and later updated in 
\cite{Martin:2009ii},
which we label as ``KMR UGDF''.
The KMR UGDF is available from the \textsc{CASCADE} Monte Carlo code 
\cite{Jung:2010si}.
The KMR UGDF (dashed lines) gives smaller cross section 
than the JH UGDF (solid lines).
The results for both UGDFs coincide for large $p_{t}$.
The larger the $f_2(1270)$ transverse momentum
the larger the range of gluon transverse momenta 
$q_{1t}$ and/or $q_{2t}$ are probed.
In the right panel of figure~\ref{fig:GJR} 
we show the $\pi\pi$-rescattering contribution.
In the $\pi\pi$-rescattering mechanism we used a L{\'e}vy parametrization of the inclusive $\pi^0$ cross section
proposed in \cite{Abelev:2012cn} for $\sqrt{s} = 7$~TeV;
see \cite{LS2020} for details.
Clearly the $\pi \pi \to f_2(1270)$
rescattering effect cannot describe the region of $p_t > 2$~GeV,
where the $gg$-fusion mechanism is a possible explanation.


\vspace{-0.4cm}
\section{Conclusions}
\vspace{-0.2cm}

In this presentation results obtained in \cite{LS2020} have been
presented.
Two different approaches for the $g^* g^* \to f_{2}(1270)$ coupling were
considered.
The coupling constants have been fixed by the Belle data
for $\gamma \gamma \to f_2(1270) \to \pi \pi$.
Then, the $g^* g^* \to f_2(1270)$ vertices have been obtained by replacing
electromagnetic coupling constant by the strong coupling constant,
modifying color factors and assuming a simple flavour structure of the $f_2(1270)$ isoscalar meson.

Our calculation of the cross section for 
$p p \to f_2(1270) + X$ were performed within the $k_t$-factorization 
approach using different unintegrated gluon distributions.

At low $f_2(1270)$ transverse momenta
the helicity-2 ($\Lambda = 2$) contribution
dominates, while the helicity-0 ($\Lambda = 0$,~T) is small.
However, the latter competes with the $\Lambda = 2$ contribution at larger transverse momenta of $f_2(1270)$. 
In the PPV formalism there are also 
$\Lambda = 0$,~L and $\Lambda = 1$
contributions which are, however, difficult to fix at present.

It has been shown that the results strongly depend on the form of
the vertex form factor which is poorly known at present. 
While one can describe the data for $p_t > 2$~GeV, it seems
impossible to describe the low-$p_t$ data.
We have shown that there one may expect $\pi \pi$ rescattering.
Adjusting one parameter of a simple model for the $\pi \pi$ rescattering 
one can describe the region of small $p_t$. 

\vspace{-0.3cm}

\acknowledgments

\vspace{-0.3cm}

The authors thank the organisers of the ICHEP 2020
conference for making this presentation of our results possible.
This work was partially supported by
the NCN Grant,
No. 2018/31/B/ST2/03537.


\vspace{-0.3cm}
\begin{thebibliography}{99}

\vspace{-0.2cm}

\bibitem{Adams:2003cc}
J.~Adams {\em et~al.}, (STAR Collaboration), 
\href{http://dx.doi.org/10.1103/PhysRevLett.92.092301}{Phys. Rev. Lett. {\bfseries 92} (2004) 092301}.

\vspace{-0.2cm}

\bibitem{Lee:thesis}
G.~R. Lee,
\newblock {Ph.D. thesis: 
\textit{Resonance production in the $\pi^+ \pi^-$ decay channel 
in proton-proton collisions at 7 TeV}, 
University of Birmingham, 2016}.
\url{http://www.hep.ph.bham.ac.uk/publications/thesis/grl_thesis.pdf}

\vspace{-0.2cm}

\bibitem{FillionGourdeau:2007ee}
F.~Fillion-Gourdeau and S.~Jeon, 
  \href{http://dx.doi.org/10.1103/PhysRevC.77.055201}{Phys. Rev. {\bfseries C77} (2008) 055201}.

\vspace{-0.2cm}

\bibitem{LS2020}
P. Lebiedowicz and A. Szczurek,
\href{https://doi.org/10.1016/j.physletb.2020.135816}{Phys. Lett. {\bf B810} (2020) 135816}.

\vspace{-0.2cm}

\bibitem{Cisek:2017gno}
A.~Cisek and A.~Szczurek, 
\href{http://dx.doi.org/10.1103/PhysRevD.97.034035}{Phys. Rev.
  {\bfseries D97} (2018) 034035}.

\vspace{-0.2cm}

\bibitem{Babiarz:2020jkh}
I.~Babiarz, R.~Pasechnik, W.~Sch{\"a}fer, and A.~Szczurek, 
  \href{http://dx.doi.org/10.1007/JHEP06(2020)101}{JHEP {\bfseries 06} (2020) 101}.

\vspace{-0.2cm}

\bibitem{Baranov:2019joi}
S.~P. Baranov and A.~V. Lipatov, 
  \href{http://dx.doi.org/10.1140/epjc/s10052-019-7134-4}{Eur. Phys. J.
  {\bfseries C79} (2019) 621}.

\vspace{-0.2cm}

\bibitem{Babiarz:2019mag}
I.~Babiarz, R.~Pasechnik, W.~Sch{\"a}fer, and A.~Szczurek, 
  \href{http://dx.doi.org/10.1007/JHEP02(2020)037}{JHEP {\bfseries 02} (2020) 037}.

\vspace{-0.2cm}

\bibitem{Lebiedowicz:2020bwo}
P.~Lebiedowicz, R.~Maciu{\l}a, and A.~Szczurek, 
  \href{http://dx.doi.org/10.1016/j.physletb.2020.135475}{Phys. Lett. B
  {\bfseries 806} (2020) 135475}.


\vspace{-0.2cm}

\bibitem{Ewerz:2013kda}
C.~Ewerz, M.~Maniatis, and O.~Nachtmann, 
  \href{http://dx.doi.org/http://dx.doi.org/10.1016/j.aop.2013.12.001}{Annals
  Phys. {\bfseries 342} (2014) 31}.



\vspace{-0.2cm}

\bibitem{Dai:2014zta}
L.-Y. Dai and M.~R. Pennington, 
\href{http://dx.doi.org/10.1103/PhysRevD.90.036004}
{Phys.Rev. D {\bfseries 90} (2014) 036004}.






\vspace{-0.2cm}

\bibitem{Pascalutsa:2012pr}
V.~Pascalutsa, V.~Pauk, and M.~Vanderhaeghen, 
  \href{http://dx.doi.org/10.1103/PhysRevD.85.116001}{Phys. Rev. {\bfseries
  D85} (2012) 116001}.

\vspace{-0.2cm}

\bibitem{Poppe:1986dq}
M.~Poppe, 
\href{http://dx.doi.org/10.1142/S0217751X8600023X}{Int. J. Mod. Phys.
  {\bfseries A1} (1986) 545}.



\vspace{-0.2cm}

\bibitem{Schuler:1997yw}
G.~A. Schuler, F.~A. Berends, and R.~van Gulik, 
  \href{http://dx.doi.org/10.1016/S0550-3213(98)00128-X}{Nucl. Phys. B
  {\bfseries 523} (1998) 423}.

\vspace{-0.2cm}

\bibitem{Shirkov:1997wi}
D.~V. Shirkov and I.~L. Solovtsov, 
\href{http://dx.doi.org/10.1103/PhysRevLett.79.1209}{Phys. Rev.
  Lett. {\bfseries 79} (1997) 1209}.

\vspace{-0.2cm}

\bibitem{Abelev:2012cn}
B.~Abelev {\em et~al.}, (ALICE Collaboration), 
  \href{http://dx.doi.org/10.1016/j.physletb.2012.09.015}{Phys. Lett. {\bfseries
  B717} (2012) 162}.





\vspace{-0.2cm}

\bibitem{Antchev:2013gaa}
G.~Antchev {\em et~al.}, (TOTEM Collaboration), 
\href{http://dx.doi.org/10.1209/0295-5075/101/21002}{EPL {\bfseries 101}
  (2013) 21002}.

\vspace{-0.2cm}

\bibitem{Aad:2014dca}
G.~Aad {\em et~al.}, (ATLAS Collaboration), 
  \href{http://dx.doi.org/10.1016/j.nuclphysb.2014.10.019}{Nucl. Phys.
  {\bfseries B889} (2014) 486}.

\vspace{-0.2cm}

\bibitem{Hautmann:2013tba}
F.~Hautmann and H.~Jung, 
  \href{http://dx.doi.org/10.1016/j.nuclphysb.2014.03.014}{Nucl. Phys.
  {\bfseries B883} (2014) 1}.




\vspace{-0.2cm}

\bibitem{Jung:2010si}
H.~Jung {\em et~al.}, 
\href{http://dx.doi.org/10.1140/epjc/s10052-010-1507-z}{Eur. Phys.
  J. {\bfseries C70} (2010) 1237}.

\vspace{-0.2cm}

\bibitem{Kimber:2001sc}
M.~A. Kimber, A.~D. Martin, and M.~G. Ryskin, 
  \href{http://dx.doi.org/10.1103/PhysRevD.63.114027}{Phys.
  Rev. {\bfseries D63} (2001) 114027}.


\vspace{-0.2cm}

\bibitem{Martin:2009ii}
A.~D. Martin, M.~G. Ryskin, and G.~Watt, 
  \href{http://dx.doi.org/10.1140/epjc/s10052-010-1242-5}{Eur. Phys. J.
  {\bfseries C66} (2010) 163}.




\vspace{-0.2cm}

\bibitem{Lebiedowicz:2016ioh}
P.~Lebiedowicz, O.~Nachtmann, and A.~Szczurek, 
\href{http://dx.doi.org/10.1103/PhysRevD.93.054015}{Phys. Rev.
  {\bfseries D93} (2016) 054015}.

\vspace{-0.2cm}

\bibitem{Lebiedowicz:2019por}
P.~Lebiedowicz, O.~Nachtmann, and A.~Szczurek, 
  \href{http://dx.doi.org/10.1103/PhysRevD.101.034008}{Phys. Rev. D {\bfseries
  101} (2020) 034008}.
  
\end{thebibliography}
\end{document}